\documentstyle[manuscript,aps]{revtex}
\begin{document}
\draft

\title{ ROLE OF FRAGMENT HIGHER STATIC DEFORMATIONS IN
THE COLD BINARY FISSION OF $^{252}$Cf}
 
\author{ A. S\u andulescu$^{1,2,3)}$, \c S. Mi\c sicu$^{1)}$, 
F. C\^ arstoiu$^{1)}$, A. Florescu$^{1,2}$,  and  W. Greiner$^{2,3)}$}

\address{ $^{1)}$ Institute of Atomic Physics, Bucharest, 
P.O.Box MG-6, Romania}
\address{ $^{2)}$ Physics Department, Vanderbilt University, Nashville, 
TN 37235, USA}
\address{ $^{3)}$ Institut f\"ur Theoretische Physik der J.W.Goethe 
Universit\"at, D-60054, Frankfurt am Main, Germany}
\maketitle

\begin{abstract} 
We study the binary cold fission of $^{252}$Cf in the frame 
of a cluster model where the fragments are born to their
respective ground states and interact via a double-folded
potential with deformation effects taken into account up to
multipolarity $\lambda=4$. The preformation factors were
neglected.
In the case when the fragments are assumed to be spherical or
with ground state quadrupole deformation, the $Q$-value
principle dictates the occurence of a narrow region around the
double magic $^{132}$Sn, like in the case of cluster radioactivity.
When the hexadecupole deformation is turned on, an entire
mass-region of cold fission in the range 138$\div$156 for the
heavy fragment arise, in agreement with the experimental observations.  
 This fact suggests that in the above mentioned mass-region, contrary
to the usual cluster radioactivity where the daughter nucleus is
always a neutron/proton (or both) closed shell or nearly closed
shell spherical nucleus, the clusterization mechanism seems to
be strongly influenced by the hexadecupole deformations rather
than the $Q$-value.  
\end{abstract}
  
\newpage

\hskip 0.25truecm { \bf 1.  Introduction }

\vskip .35truecm

In the binary nuclear fission of actinide nuclei the fragments are
usually formed in highly-excited states which subsequently decay
to their ground-states by emitting neutrons and gamma rays. 
However a small fraction of these fragmentations will attain a 
very high kinetic energy $TKE$ which is very close to the 
corresponding binary decay energy $Q$. Since in this case the 
fragments are formed with excitations energies close to their 
ground-states no neutrons are emitted. 
Milton and Fraser \cite{Mil61} were the first who noticed that some
of the fission fragments are produced at such high kinetic
energies that the emerging nuclei are formed nearly in their ground-state. 
Later on Signarbeux et al. \cite{GUE78} confirmed the
previous interpretation by determining the mass distributions
of the primary fragments for the highest values of the kinetic energy.
They concluded that even before the scission takes place we
deal with a superposition of two fragments in their ground
state, from which the {\em cold fragmentation} term emerged. 
An interesting remark they made was that the odd-even
fluctuations of $Q$ due to nucleon pairing were not present also
in the $TKE_{max}$ values. In their view this smoothing of the
odd-even effect was a consequence of a pair-broken from one of
the fragments. 
The probability for neutronless fission is 0.0021$\pm$0.0008 for
$^{252}$Cf. 

In the last years the cold (neutronless) fission of many actinide nuclei 
into fragments with masses from $\approx$70 to $\approx$ 160 was an
intensivelly studied phenomenon 
\cite{GB91,HKB93,Ben93,Sch94,Ham94,Ter94,San96}. 
An important step in the understanding of the cold fission phenomenon 
was the observation that the final nuclei are generated in
their ground states or some low excited states, which prompted some
authors to relate these decays to the spontaneous emission  
of light nuclei (cluster radioactivity) such as alpha particles and 
heavier clusters ranging from $^{14}$C to $^{34}$Si \cite{SG92}.

The fragments emitted in binary cold decays are produced with
very low or even zero internal excitation energy and consequently with 
very high kinetic energy $TKE = Q - TXE$.  
In order to achieve such large $TKE$ values, both fragments
should have very compact shapes at the scission point and 
deformations close to those of their ground states \cite{GB91,SFG89}.

The first direct observation of cold (neutronless) binary
fragmentations in the spontaneous fission of $^{252}$Cf was
made by using the multiple Ge-detector Compact Ball
facility at Oak Ridge National Laboratory \cite{Ham94,Ter94}, 
and more recently with the Gammasphere consisting of 72 detectors 
\cite{San96}. Using the triple-gamma coincidence technique, the 
correlations between the two fragments were observed unambiguously.

In these cold fragmentations, some indications of a third light 
fragment such as  $\alpha$, $^{6}$He and $^{10}$Be clusters
\cite{Ram96,San97,Ram97}, were also reported. 

In a recent series of publications \cite{MCGP96,Gon97} the group of
T\"ubingen reported some interesting results on the spontaneous decay 
of $^{252}$Cf using a twin ionization chamber.
Two distinct mass regions of cold fission were observed : the first  
extending from the mass split 96/156 up to 114/138 and the second one   
comprising only a narrow mass range around the mass split 120/132. 

In the present paper, based on a  cluster model similar to the 
cluster model used for cluster radioactivity, we estimated the 
relative isotopic yields for the spontaneous cold 
binary fission of $^{252}$Cf. These isotopic yields
are given by the ratio of the penetrability through the
potential barrier between the two final 
fragments for a given mass and charge splitting, over
the sum of penetrabilities for all possible 
fragmentations. 

The corresponding barriers were evaluated
using the double folding potential with M3Y nucleon-nucleon
effective interactions and realistic ground state deformations 
including the  octupole and hexadecupole ones \cite{San97}.

We were mainly concerned with the study of the influence 
of the fragment deformations on the yields and we concluded that the 
occurence of the two mass-regions of cold fission is determined 
essentially by the ground state hexadecupole deformations.
\vskip 0.5truecm

\hskip 0.25truecm  { \bf 2. Deformation Dependent Cluster Model }

\vskip .25truecm

  In the present paper we consider a deformation dependent cluster model,
similar to the one-body model used for the description of cluster
radioactivity \cite{SG92}. The initial nucleus is assumed to be
already separated into two parts, a heavy one and a light cluster, 
and the preformation factors for the fragments are not taken into 
account. An advantage of this model is that the barrier between the 
two fragments can be calculated quite accurately due to the fact
that the touching configurations are situated inside of the barriers.
The $Q$ values and the deformation parameters contain all
nuclear shell and pairing effects of the corresponding
fragments. 

The barriers were calculated using the double folding model for heavy 
ion interaction 
 $$ V_{F} ({\bf R}) = \int d{\bf r}_{1} d{\bf r}_{2}~
\rho_{1} ({\bf r}_{1}) \rho_{2} ({\bf r}_{2}) v({\bf s}) \eqno(1) $$
where $\rho_{1(2)}({\bf r})$ are the ground state one-body densities 
of the fragments and $v$ is the $NN$ 
effective interaction. The separation distance between two interacting 
nucleons is denoted by ${\bf s}={\bf r}_{1}+{\bf R}-{\bf r}_{2}$, where
$R$ is the distance between the c.m. of the two fragments.
We have choosen the $G$-matrix M3Y effective interaction 
which is representative for the so called local and density independent 
effective interactions \cite{BS97}. This interaction is 
particularly simple to use in folding models since it is parametrized  
as a sum of 3 Yukawa functions in each spin-isospin $(S,T)$ channel.
In the present study the spin and spin-isospin dependent components have 
been neglected since for a lot of fragments involved in the calculation 
the ground state spins are unknown. 
The spin-spin component of the heavy-ion potential can be neglected 
here since it is of the order ${1\over {A_{1}A_{2}}}$.
Only the isoscalar and isovector components have been retained in the 
present study for the central heavy ion interaction.

The M3Y interaction is dominated by the one-nucleon knock-on exchange 
term, which leads to a nonlocal kernel. In the present 
case the nonlocal potential is reduced to a zero range pseudopotential 
$\hat{J}_{00} \delta({\bf s})$, with a strength depending slightly on the 
energy. We have used the common prescription \cite{BS97}
$\hat{J}_{00}$ = -276~MeV$\cdot$fm$^{3}$
neglecting completely the small energy dependence. For example, the 
odd-even staggering in the $Q$-value for a fragmentation channel, which 
is tipically of the order $\Delta Q$=2 MeV, leads to a variation with 
$\Delta \hat{J}_{00}$=-0.005${\Delta Q}\over\mu$ MeV$\cdot$fm$^{3}$
with $\mu\approx$100.
The one-body densities in (1) are taken as Fermi distributions in the 
intrinsic frame
$$\rho({\bf r}) =\frac{\rho_0}{1+e^{\frac{r-c}{a}}}\eqno(2)  $$ 
with 
$c=c_{0}(1+\sum_{\lambda\geq 2}\beta_{\lambda}Y_{\lambda 0}(\Omega))$.
Only static axial symmetric deformations are considered. 
The half radius $c_{0}$ and the diffusivity $a$ are taken from the
liquid drop model \cite{Mol95}.
The normalization constant $\rho_0$ is determined by requiring the 
particle number conservation
$$\int r^{2}dr~d\Omega\rho(r,\Omega) = A\eqno(3) $$
and then the multipoles are computed numerically
$$ \rho_{\lambda}(r) = \int d\Omega\rho(r,\Omega)Y_{\lambda 0}(\Omega).
\eqno(4)$$
Once the multipole expansion of the density is obtained, the integral in 
(1) becomes 
$$
V_{F}({\bf R},\omega_{1},\omega_{2}) 
= \sum_{\lambda_{1}\mu_{1}\lambda_{2}\mu_{2}} 
D_{\mu_{1}0}^{\lambda_{1}}(\omega_{1})
D_{\mu_{2}0}^{\lambda_{2}}(\omega_{2})
I_{\lambda_{1}\mu_{1}\lambda_{2}\mu_{2}} \eqno(5)$$
where \cite{CL92}
$$I_{\lambda_{1}\mu_{1}\lambda_{2}\mu_{2}} = 
\sum_{\lambda_{3}\mu_{3}}B_{\lambda_{1}\mu_{1}\lambda_{2}\mu_{2}}^
{\lambda_{3}\mu_{3}}
\int r_{1}^{2}dr_{1} r_{2}^{2}dr_{2} 
\rho_{\lambda_{1}}(r_{1}) \rho_{\lambda_{2}}(r_{2})
F_{\lambda_{1}\lambda_{2}\lambda_{3}}^{v}(r_{1},r_{2},R)
\eqno(6)$$
and 
$$ 
F_{\lambda_{1}\lambda_{2}\lambda_{3}}^{v}(r_{1},r_{2},R) = 
\int q^{2}dq{\tilde v}(q)
j_{\lambda_{1}}(qr_{1})j_{\lambda_{2}}(qr_{2})j_{\lambda_{3}}(qr_{3}).
\eqno(7)$$
Above, $D_{\mu 0}^{\lambda}(\omega)$ stands for the Wigner rotation matrix 
describing the orientation $\omega$ of the intrinsic symmetry axis with 
respect to the fixed frame, ${\tilde v}(q)$ denotes the Fourier transform 
of the interaction and $j_{\lambda}$ are the spherical Bessel functions. 
The matrix $B$ in (6) is defined in \cite{CL92} and contains
selection rules for coupling angular momenta. Only 
$\lambda_{1}+\lambda_{2}+\lambda_{3}= $even, are allowed.
When $\beta_{\lambda}\neq 0$, $\lambda=2,3,4$ for both fragments, 
the sum in (5) involves 36 terms for a nose-to-nose configuration with
$\lambda_{3}\le 6$.   
For most of the fragmentation channels studied 
here, large quadrupole, hexadecupole, and occasionally octupole 
deformations are involved. Therefore a Taylor expansion method for 
obtaining the density multipoles turns out to be unsuitable. 
On the other hand, a large quadrupole deformation induces according to 
(4) nonvanishing 
multipoles with $\lambda$=4 and 6 even if $\beta_4$=$\beta_6$=0. 
Therefore for a correct calculation of 
(4), a numerical method with a truncation error of order O$(h^7)$ is 
needed in order to ensure the orthogonality of spherical harmonics with 
$\lambda\le 6$.
Performing the integrals (6) and (7) we have used a numerical method with 
a truncation error of the order O$(h^9)$. All short range wavelength 
($q\le 10$~fm$^{-1}$) have been included and particular care has been taken 
to ensure the convergence of the integrals with respect to the 
integration step and the range of integration.

At the scission configuration two coaxial deformed fragments in 
contact at their tips were assumed. For quadrupole deformations we 
choose two coaxial prolate spheroids due to the fact that the prolate 
shapes are favoured in fission. 
It is known that for each oblate minimum always corresponds another 
prolate minimum. For pear shapes, i.e. fragments with quadrupole and 
octupole deformations, we choose opposite signs for the octupole 
deformations, i.e. nose-to-nose configurations (see Fig.1). 
For hexadecupole deformations we choose only positive 
signature, because it leads to a lowering of the barriers in comparison 
with negative ones and consequently they are much more favoured in 
fission (see Fig.2).

 In order to ilustrate the influence of deformations on the barriers 
we displayed in Fig.3  the M3Y-folding 
multipoles for $^{106}$Mo and $^{146}$Ba with all deformations 
included. 
The octupole component is large in the interior but gives negligible
contribution in the barrier region in contrast to the hexadecupole one.
Next, in Fig.4 we are illustrating for the same partners the
cumulative effect of high rank multipoles on the barrier. 
\vskip 0.5truecm

\hskip 0.25truecm  { \bf 3. Cold Fission Binary Isotopic Yields}
 
\vskip .25truecm

   We should like to stress again that in our simple cluster
model the preformation factors for different channels are neglected, 
i.e. we use the same assault frequency factor $\nu$  for the
collisions with the fission barrier for all fragmentations.
It is generally known that the general trends in alpha decay of heavy nuclei
are very well described by barrier penetrabilities, the preformation
factors becoming increasingly important only in the vicinity of
the double magic nucleus $^{208}$Pb. On the other hand
the cold binary fragmentation of $^{252}$Cf was also reasonably
well described using constant preformation factors
\cite{San96,FSCG93}. However in this case too, as we shall see
later, around the double-magic nucleus $^{132}$Sn the preformation
turn out to be of capital importance. 
Eventually, as the experimental data become more accurate we would
be able to extract some fragment preformation factors
and discuss the related nuclear structure effects.
 
In the laboratory frame of reference the $z$-axis was taken as
the initial fissioning axis of the two fragments,
with the origin at their point of contact. 
The potential barriers $V_{F} - Q_{LH}$ between the two
fragments are high but rather thin with a width
of about 2 to 3 fm. As an illustration, we show in Fig.5
a typical barrier between $^{146}$Ba and $^{106}$Mo,
as a function of the distance $R_{LH}$ between their 
center of mass. Here $Q_{LH}$ is the decay
energy for the binary fragmentation of $^{252}$Cf.

For the two fragments, the exit point from their potential barrier 
is at $R_{LH}$ typically between 16 and 17 fm (see Fig.5) which
supports our cluster model. 

The penetrabilities through the double-folded potential barrier 
between the two fragments were calculated by using the WKB
approximation 

  $$P = \exp \left\lbrace -{2 \over \hbar} \int_{s_{i}}^{s_{o}} 
  \sqrt{~ 2 \mu~ [~V_{F}(s)-Q_{LH}~]~}~~ds \right\rbrace   
\eqno(8) $$
  where $s$ is the relative distance,
 $\mu$ is the reduced mass and $s_{i}$ and $s_{o}$ 
 are the inner and outer turning points, defined
 by $V_{F}( s_{i} ) = V_{F}( s_{o} ) = Q_{LH} $.

The barriers were computed with the LDM parameters $a_p=a_n$=0.5~fm, 
$r_{0p}=r_{0n}=(R-{1\over R})A^{-1/3}~$fm with 
$R=1.28A^{1/3}+0.8A^{-1/3}-0.76$.

Accurate knowledge of $Q$ values is crucial for the calculation,
since the WKB penetrabilities are very sensitive to them.
We obtained the $Q$ values from experimental mass tables 
\cite{WAH88}, and for only a few of the fragmentations the
nuclear masses were taken from the extended tables of M\"oller et
al. \cite{Mol95} computed using a macroscopic-microscopic model.

Let us consider for the beginning only the relative isotopic
yields corresponding to true cold (neutronless) binary fragmentations 
in which all final nuclei are left in their ground state.
These relative isotopic yields are given by the expression
($A_{1}=A_{L},A_{2}=A_{H}$)
 $$ Y ( A_{1}, Z_{1} ) = { P ( A_{1}, Z_{1} ) \over \sum_{A_{1} Z_{1}}
  P ( A_{1}, Z_{1} ) } ~~\cdot \eqno(9)  $$
As we mentioned above the fragment deformations were choosed to be the
ground state deformations of M\"oller et al.\cite{Mol95}, computed
in the frame of the macroscopic-microscopic model. 
In Fig.6 we represented separately these deformations for the light 
$A_{L}$ and heavy $A_{H}$ fragments for odd and even charge $Z$. We
can see that the light fragments, have mainly quadrupole deformations
in contrast to the heavy fragments, which have all types of deformations. 
The octupole deformations are non-zero in a small heavy fragment mass 
number region 141 $\leq A_{H}\leq 148$. The fragments with mass number 
$A_{L}\leq 92$ and $A_{H}\leq 138$  are practically spherical.

The computed M3Y-fission barriers heights, for different assumptions: 
no deformations,
including the quadrupole ones, including the quadrupole and octupole
ones and for all deformations, together with the corresponding 
$Q$-values are represented in Fig.7 for odd $Z$ and even $Z$
separetely. We notice the large influence due to the
quadrupole deformations but also the hexadecupole ones are
lowering the barriers very much.  
The octupole deformations in the mass region  
$141\leq A_{H}\leq 148$ have a smaller effect as we expected. This
is a illustration of the difference between
cluster radioactivity, which is due only to the large $Q$-values and
the cold fission which is due mainly to the lowering of the barriers
due to the fragment deformations. Both processes are cold
fragmentation phenomena.

The computed yields in percents, for the splittings represented by
their fragment deformation parameters in Fig.6 or by their
barrier heights in Fig.7, are given in Fig.8 for
spherical fragments ($\beta_{i}=0$), for quadrupole deformations 
($\beta_{2}$) and for all deformations 
($\beta_{2}+\beta_{3}+\beta_{4}$) at zero excitation energy. 
We can see that when the fragments are assumed to be spherical the 
splittings with the highest $Q$-values, which correspond to real spherical 
heavy fragments(see Fig.6), i.e. for charge combinations $Z_1/Z_2=$ 48/50, 
47/51 and 46/52 are the predominant ones. 
As we mentioned before this situation is similar 
with the cluster radioactivity were the governing principle is the 
$Q$-value. Due to the staggering of $Q$-values (see Fig.7) the
highest yields are for even-even splittings. 
By including the $\beta_2$ deformations few asymmetric splittings 
exists.
For all deformations more asymmetric yields appear. Now the principal 
yields are for $Z_{1}/Z_{2}$= 38/60, 40/58, 41/57 and 42/56
along with 44/54, 46/52 and 47/51.
This is due to the fact that the influence of the fragment 
deformations on the yields overcome the influence of $Q$-values in the more 
asymmetric region. This illustrate the fact that cold fission is a cold
rearrangement process in which all deformations are playing the main
role and not the  $Q$-values. The staggering for odd $Z$ fragmentations 
like $Z_{1}/Z_{2}$ = 39/59, 41/57, 43/55, 45/53 and 47/51 or odd $N$
fragmentations is recognized at first glance. 
However, by the introduction of the density levels this staggering
is reversed. The largest yields will be for odd $Z$ and/or 
$N$ fragmentations.

In the next figure we represented the mass yields 
$Y_{A_{2}}=\sum_{Z_{2}}Y(A_{2},Z_{2})$ (Fig.9) for spherical fragments 
($\beta_i$=0), for quadrupole deformations ($\beta_2\neq$0) and for
all deformations ($\beta_i\neq$0). We can see in the spherical case that 
the main mass yields are centered around $A_{2}$=132. All these 
heavy fragments are spherical or nearly spherical 
(with a small prolate deformation) and have high-$Q$ values. 
Since other spherical fragments does not arise in the yields diagram  
it occurs that in the spherical case the $Q$-value is the dictating 
principle. When we turn on the quadrupole deformation a rearrangent in this
spherical region takes place. The yield corresponding to $A_{2}$=132 is 
still important but the one for $A_{2}$=134 takes over although the 
maximum decay energy of the first mass split $Q_{max}$ is larger than that  
of the former. In this case the larger quadrupole deformation of the light 
partner decides the augmentation of the $A_{2}$=134 yield.
When we include the higher multipole deformations, i.e. octupole and 
hexadecupole deformations the yields diagram will change drastically 
over the whole mass range. 
First of all, in the spherical region the mass-splitings yields 
$A_{2}$=132, 134 are lowered whereas their odd neighbours are 
augmented. Once again this is a consequence of the fact that the 
hexadecupole deformations of the odd light partners are slightly larger. 
But the most important change occurs in the mass region 
$A_{2}$=138$\div$156 where a whole bunch of splittings show up with yields 
greater than 0.01$\%$. This is, beyond any doubt, an effect due to the 
hexadecupole deformations. As can be infered from Fig.6 the above mentioned  
mass region is characterized by noticeable values of the hexadecupole
deformation. Before adding the hexadecupole deformation this region was
completely desertic whereas after the inclusion of $\beta_4$  the most 
pronounced peaks are $A_2$= 138, 140, 146, 150 and 154. It is the place 
to mention that the first mass region, in the cold fission of $^{252}$Cf 
reported in the paper of G\"onnenwein et al.\cite{Gon97} coincides with the 
range obtained by us employing a deformation dependent cluster model.
However in order to reproduce completely the experimental data 
we have to underline the elements that have to be supplied further in our
model. 
First, in the spherical region, the experiment claim a mass region of 
cold fission centered around $A_2$=132, instead of $A_2$=134 as we 
obtained. However this misfit was to be expected since as we mentioned in   
the beginning of our paper we didn't included the preformation factors.
In the case of the doubly magic nucleus $^{132}$Sn this assumption 
proves to be unsatisfactory. As has been advocated by the T\"ubingen group
\cite{Gon97} this is a possible manifestation of heavy-cluster decay. 
Therefore it is very likely that in this case the preformation factor, 
which multiplies the penetrability, is larger than for the neighbouring 
nuclei, which could then account for the discrepancies between our 
calculations and experimental data. 
However an encouraging experimental point which supports 
our calculations is the fact that the even masses 134 and 136 
are accompanying the leading yield for 132. In fig.10 we compare the   
total yields for 132 (left side) and 134 (right side). We see that the 
$Z$-splitting  corresponding to the spherical $^{134}$Te dominates in 
all the three cases, because, as we mentioned earlier its light partner has
a sensitive quadrupole deformation and a non-vanishing hexadecupole one. 
Its $Z$ partner $^{134}$Sn has a smaller hexadecupole deformation.
The same reasoning apply to $A_2$=136. Therefore it could be possible 
that in the case of these nuclei the deformation dictates the yield 
magnitude rather than the magic number in protons or neutrons. 
The experimental determination of the double fine structure in this 
region will, hopefully, clarify the situation.

The {\em hexadecupole deformed} region, extending from 138 to 156,  
obtained in the frame of our cluster model, presents also some  
discrepancies compared to the experimental findings. The main problem  
that we faced here concerns the odd-even effect which seems to be very 
strong in this region according to the T\"ubingen group \cite{MCGP96}. 
The things can be understood as follows: 
In the vicinity of the ground state, the level densities of odd mass nuclei  
are much larger than for even nuclei and consequently it will be more 
probable to observe cold fission for odd-odd mass splits in comparison 
to even-even mass splits. Since in our present calculations the level 
density of fragments is not taken into account our results points to an 
enhancement of even-even mass splits with respect to the odd-odd mass 
splits. In a preceding paper \cite{San97} the effect of level density was 
incorporated in the calculation of yields by means of the Fermi Back-shifted  
Model valide also for small excitation energies.  
In order to get a rough idea of how the odd-even effect influence  
the yields, we simply shift the decay energy by the fictious ground-state 
position $\Delta$ taken from the global analysis of Dilg et al. 
\cite{DSVU73}, $Q^{*}=Q-\Delta$. In fig.11 we represented the same thing  
like in Fig.10 but with the above mentioned shift in the $Q$-value.
It is obvious from the inspection of this figure that except 
$A_{2}$=138 , the odd splittings take over, in agreement with the 
experimental data.
It is worthwile to stress once again that in our view the mass region 
extending  from 138 to 156 the hexadecupole deformation is the leading 
mechanism responsible for the cold fragmentation of $^{252}$Cf. 
The lowering of the barriers due to hexadecupole deformation increase 
dramatically the penetrabilities and eventually the yields.
In figure 12 we represented the yields for the $Z$-splittings of
$A_2$=143. Comparing the first two cases we see that the yields
are almost unsensitive to quadrupole deformation. When the
hexadecupole deformation is included the distribution changes, all
the yields being shifted uniformly (in the log scale) towards 
magnitudes four times larger.
It is worthwile to notice before ending this section that the
octupole deformations are not inducing the tremendous changes
that the hexadecupole does.
\vskip .5truecm

\hskip .75truecm { \bf 4. Discussions and Conclusions }

\vskip .25truecm

The deformation dependent cluster model which we used in this paper for 
calculating the isotopic yields associated to cold binary fission, 
predicts a large number of favored binary splittings in which one or both 
fragments are well deformed in their ground states.
For cold binary fission the initial scission configurations are known : 
the fragment deformations should be essentially those of the
ground state deformations.

The main result obtained in our paper represents the theoretical 
confirmation of the existence of two distinct regions of
$^{252}$Cf cold fission. The results indicate two different mechanisms.
In the heavy mass region situated between 138 and 156, the hexadecupole
deformation gives rise to a large number of splittings. Here the
shell closure in neutrons or protons seems to not be involved.
Although the shell effects should play an important role in the
odd-even differences by enhancing the odd-odd mass splits with
respect to the even-even one, our result emphasize that the
fragments are emitted with the deformations corresponding to
those of the ground state. 
In the spherical region our results give only a hint of the 
importance of the magic nucleus $^{132}$Sn which is susceptible to be produced in
a heavy clusterization process, similar to that for light
clusters \cite{SG77}.
 Here the decay mechanism should be similar to the light cluster
radioactivity, the daughter nucleus $^{132}$Sn being traded for 
$^{208}$Pb and the heavy cluster $^{120}$Cd for $^{14}$C.

The results reported in this paper are pointing to the
importance of deformations included in the cold fission model
since the $Q$-value seems to be no longer the absolute ruler of
the process like in the case  of cluster radioactivity. 

In the future the investigations should be extended in such a
way to explain also the yields structure at finite excitation energy.
\vskip 0.5truecm

\vfill\eject

\vfill \eject

\vskip 7truecm
\centerline {\bf Figure  Captions }
\vskip 1truecm

 ${\bf Fig.~1.}$ Density plots of $^{106}$Mo
and $^{146}$Ba fragments, placed at $R$=15 fm, considered with quadrupole
and octupole deformations. In the upper part are represented the 
prolate-prolate, oblate-prolate positions and in the lower part two pear shapes 
nose to back and nose to nose. The positions are given by the deformation 
signs.
 \vskip 1.0truecm

 ${\bf Fig.~2.}$ Same as for Fig.1. The influence of different signs of 
hexadecupole deformations on $^{106}$Mo and $^{146}$Ba densities in the 
presence of large quadrupole and octupole deformations. The penetrability 
is maximized for $\beta_{4}>$0 configurations.

\vskip 1.0truecm

 ${\bf Fig.~3.}$ The influence of the M3Y-folding multipoles on the 
barrier between $^{106}$Mo and $^{146}$Ba. Notice that the main effect 
is due to $\lambda_{3}=2$. The influence of $\lambda_{3}=3$ is large 
but less important in the barrier region compared with the induced 
deformations $\lambda_{3}=5$ and $\lambda_{3}=6$ 
\vskip 1.0truecm

${\bf Fig.~4.}$ The cumulative effect of high rank multipoles on the 
barrier between $^{106}$Mo and $^{146}$Ba. We considered the  
deformations $\beta_{3}$ and $\beta_{4}$ 
much larger than the real ones in order to 
illustrate the effect of deformations.
\vskip 1.0truecm

${\bf Fig.~5.}$ The barrier between $^{146}$Ba and $^{106}$Mo as a
function of the distance $R_{HL}$ between their centers of mass. 
By $Q_{LH}$ we denote the decay energy.
\vskip 1.0truecm

${\bf Fig.~6.}$ The assumed $\beta_{2}$, $\beta_{3}$, $\beta_{4}$
ground state fragment deformations \cite{Mol95}. 
We can see that the light fragments 
$(Z_1,A_1)$ have mainly quadrupole deformations in contrast to the 
heavy fragments $(Z_2,A_2)$. The octupole deformations are existing in a 
small mass region 141$\leq A_2\leq$148 whereas the hexadecupole 
deformations are important in the region 138$\leq A_2\leq$158. 
The fragments with masses $A_1\leq$94 and $A_2\leq$138 
are practically spherical.
\vskip 1.0truecm

${\bf Fig.~7.}$ The barrier heights for all considered fragmentations  
channels represented 
for different charges $Z_1$ and mass numbers $A_1$ of the light fragment.
\vskip 1.0truecm

${\bf Fig.~8.}$ The true cold fission yields in percents for all 
fragmentations channels computed with the LDM 
parameters, for spherical nuclei, with the inclusion of quadrupole 
deformations and with all deformations at zero excitation energy.

\vskip 1.0truecm

${\bf Fig.~9.}$ The mass yields $Y_{A_2}=\sum_{Z_2}Y(A_2,Z_2)$ in 
percents, as a function of light fragment mass computed with LDM parameters.
Calculations without deformations ($\beta_{2,3,4}$=0) enhance 
only the spherical region $ A_2 \leq$ 136; the inclusion of 
quadrupole deformations ($\beta_2\neq$0) enhances the yield with 
$A_2 =$134; for all deformations there are two main mass yields regions,  
i.e. 133$\leq A_2 \leq$136 and 138$\leq A_2 \leq$156.

\vskip 1.0truecm

${\bf Fig.~10.}$ 
The yields for the $Z$-splittings of $A_2$=132, 134 in percents 
computed with LDM-parameters. 

\vskip 1.0truecm

${\bf Fig.~11.}$ The mass yields $Y_{A_2}=\sum_{Z_2}Y(A_2,Z_2)$ in 
percents, as a function of light fragment mass computed with LDM parameters 
with the decay energy modified $Q^{*}=Q-\Delta$.
The odd-odd mass splitings are this time favoured.

\vskip 1.0truecm

${\bf Fig.~12.}$ 
The yields for the $Z$-splittings of $A_2$=143 in percents 
computed with LDM-parameters. Calculations without deformations and with
the inclusion of quadrupole deformation  give nearly the same yields.
The inclusion of hexadecupole deformation increase uniformly by 4 orders 
of magnitude the yields.

\vfill \eject

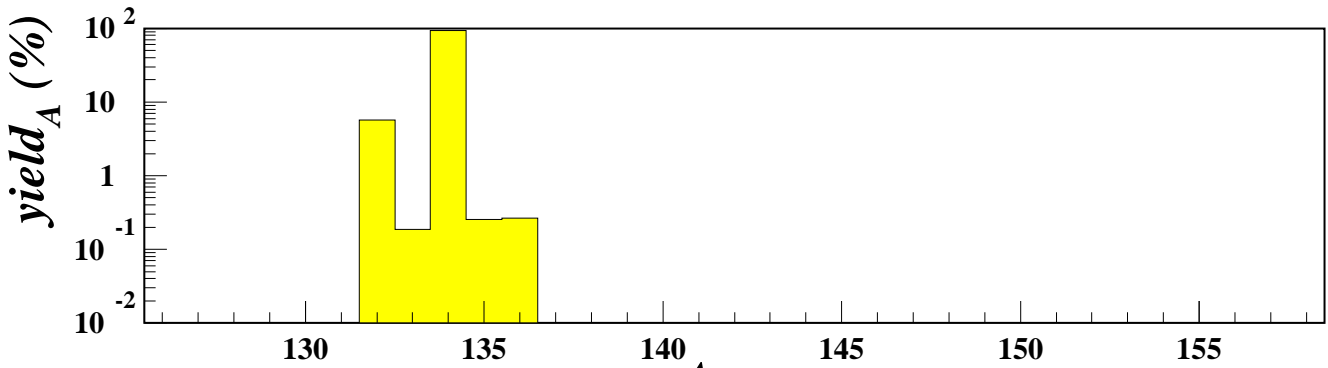
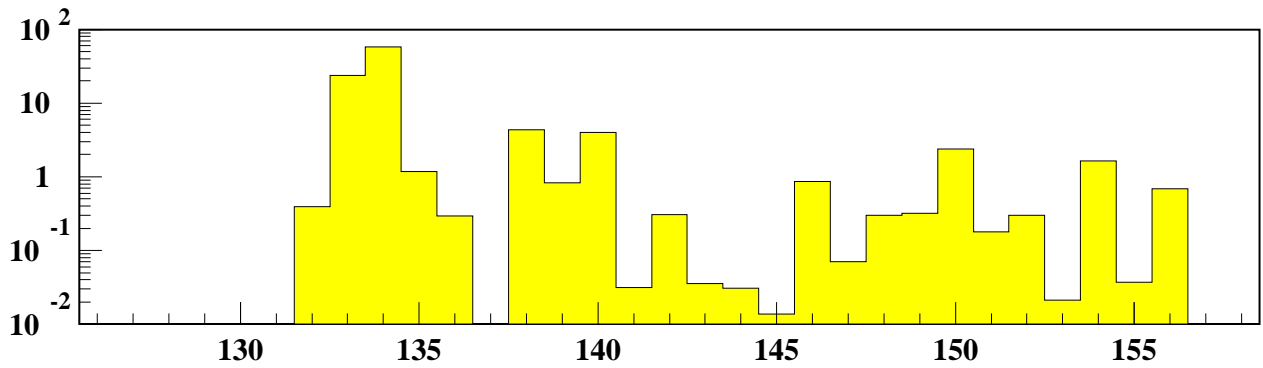
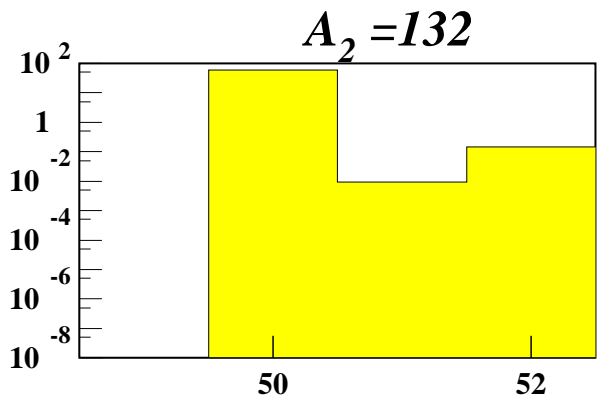
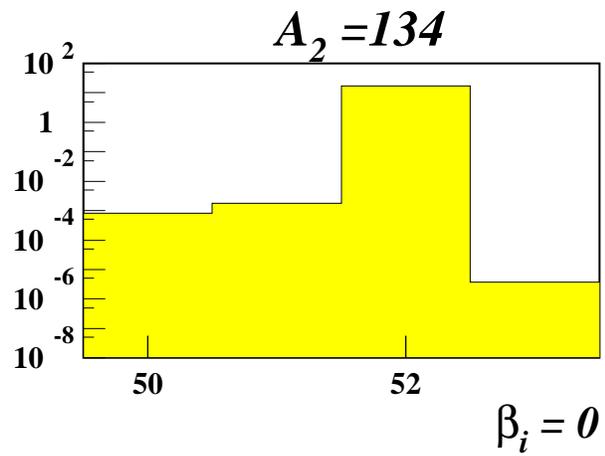
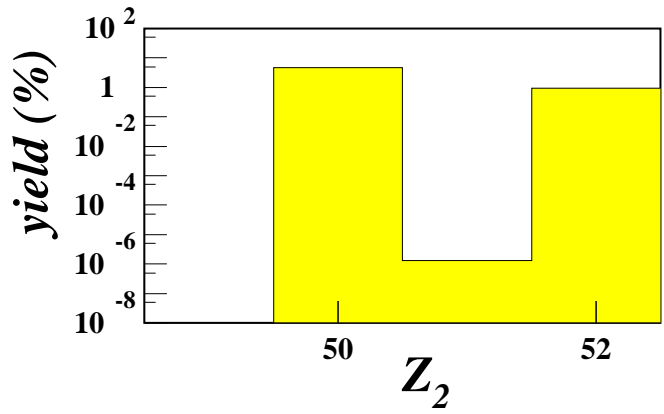
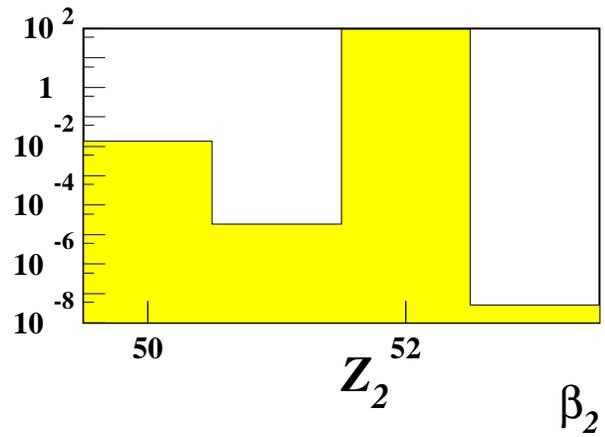
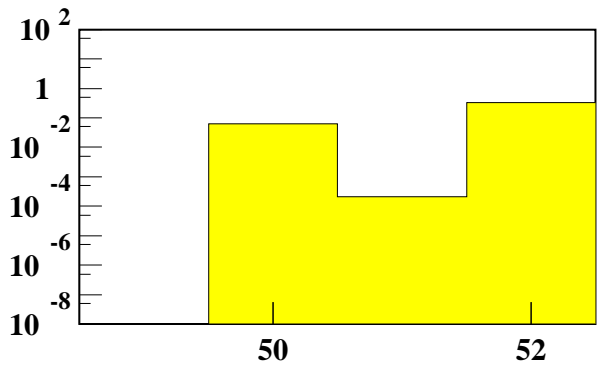
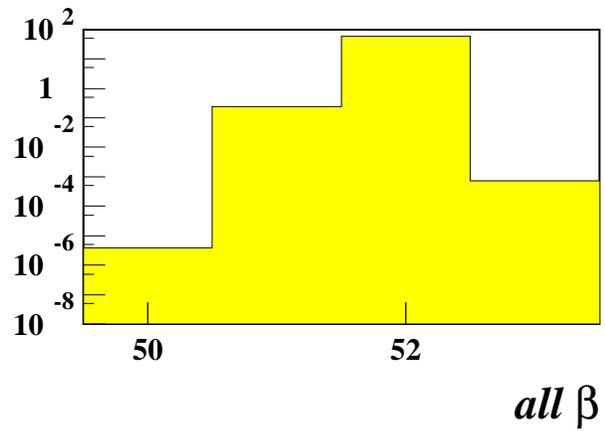
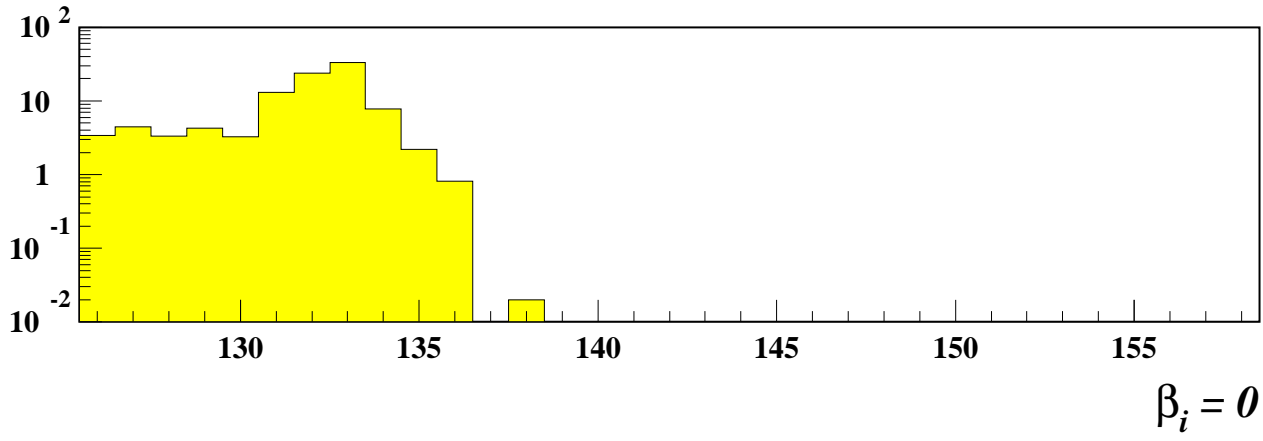
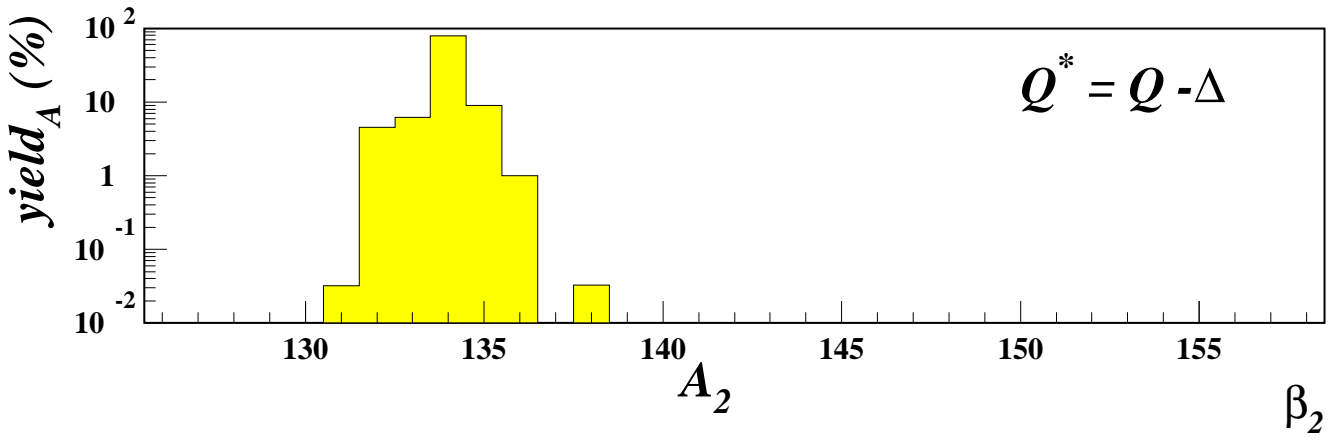
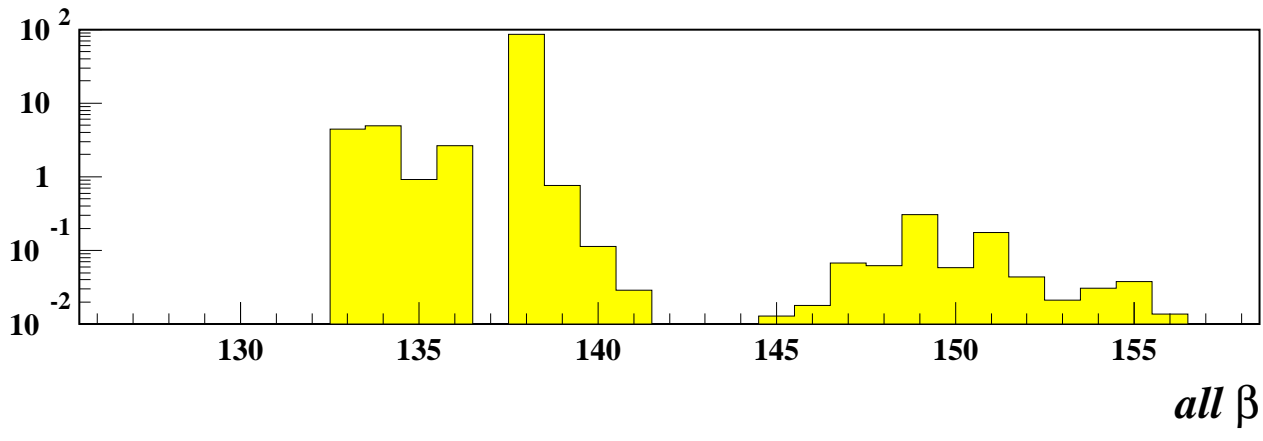
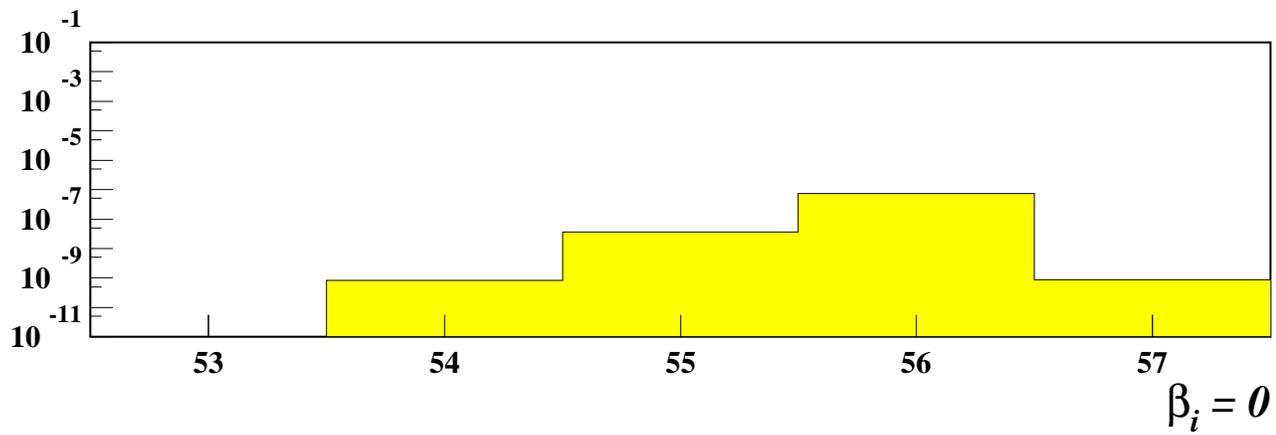
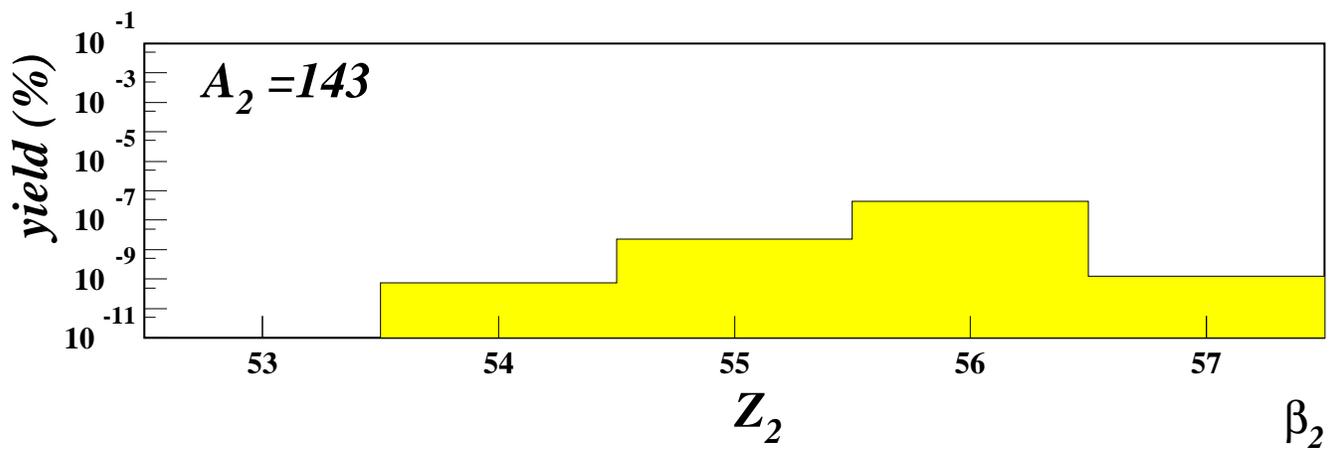
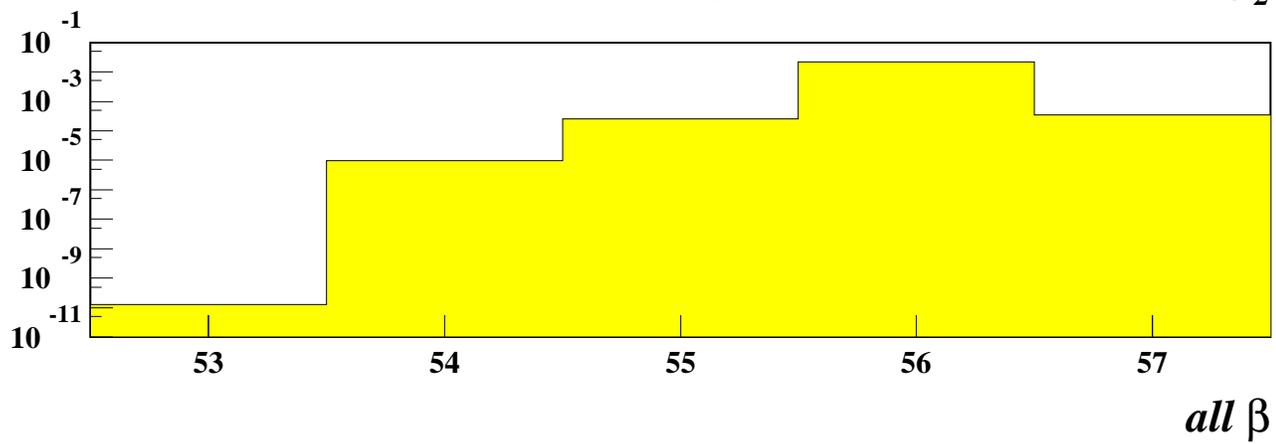
\end{document}